\documentclass[sigconf]{acmart}
\AtBeginDocument{%
  }

\copyrightyear{2026}
\acmYear{2026}
\setcopyright{cc}
\setcctype{by}
\acmConference[WSDM Companion '26]{The Nineteenth ACM International Conference on Web Search and Data Mining}{February 22--26, 2026}{Boise, ID, USA}
\acmBooktitle{The Nineteenth ACM International Conference on Web Search and Data Mining (WSDM Companion '26), February 22--26, 2026, Boise, ID, USA}
\acmDOI{10.1145/3779211.3795739}
\acmISBN{979-8-4007-2358-2/2026/02}

\usepackage{enumitem}
\usepackage{tabularx,array,makecell,booktabs,multirow}
\newcolumntype{Y}{>{\raggedright\arraybackslash}X}
\newcolumntype{C}{>{\centering\arraybackslash}X}
\newcolumntype{L}[1]{>{\raggedright\arraybackslash}p{#1}}
\usepackage{tikz}
\usetikzlibrary{shapes,arrows.meta,positioning,fit,backgrounds,calc,decorations.pathreplacing,shadows,patterns}
\begin{document}

\title{Multi-Agent Video Recommenders:\\ Evolution, Patterns, and Open Challenges}


\author{Srivaths Ranganathan}
\orcid{0009-0008-1358-2974}
\affiliation{%
  \institution{Google LLC}
  \city{Mountain View}
  \country{USA}}
\email{srivaths@google.com}

\author{Abhishek Dharmaratnakar}
\orcid{0009-0002-7335-8233}
\affiliation{%
  \institution{Google LLC}
  \city{San Bruno}
  \country{USA}}
\email{dharmaratnakar@google.com}

\author{Anushree Sinha}
\orcid{0009-0008-3189-6707}
\affiliation{%
  \institution{Google LLC}
  \city{Mountain View}
  \country{USA}}
\email{sinhaanushree@google.com}

\author{Debanshu Das}
\orcid{0009-0005-0233-4623}
\affiliation{%
  \institution{Google LLC}
  \city{Mountain View}
  \country{USA}}
\email{debanshu@google.com}

\renewcommand{\shortauthors}{Ranganathan et al.}

\begin{abstract}
 Video recommender systems are among the most popular and impactful applications of AI, shaping content consumption and influencing culture for billions of users. Traditional single-model recommenders, which optimize static engagement metrics, are increasingly limited in addressing the dynamic requirements of modern platforms. In response, multi-agent architectures are redefining how video recommender systems serve, learn, and adapt to both users and datasets. These agent-based systems coordinate specialized agents responsible for video understanding, reasoning, memory, and feedback, to provide precise, explainable recommendations. 

In this survey, we trace the evolution of multi-agent video recommendation systems (MAVRS). We combine ideas from multi-agent recommender systems, foundation models, and conversational AI, culminating in the emerging field of large language model (LLM)-powered MAVRS. We present a taxonomy of collaborative patterns and analyze coordination mechanisms across diverse video domains, ranging from short-form clips to educational platforms. We discuss representative frameworks, including early multi-agent reinforcement learning (MARL) systems such as MMRF and recent LLM-driven architectures like MACRec and Agent4Rec, to illustrate these patterns. We also outline open challenges in scalability, multimodal understanding, incentive alignment, and identify research directions such as hybrid reinforcement learning–LLM systems, lifelong personalization and self-improving recommender systems. 
\end{abstract}



\maketitle

\section{Introduction and Motivation}

Recommender systems (RSs) have become essential for navigating the
vast and growing landscape of video on the internet
\cite{liebman2015djmc,adomavicius2005,ricci2011recsys}.  
They curate personalized feeds, improve user satisfaction, and support
the attention economy across platforms for short-form entertainment,
music streaming, live broadcasts, and educational media. Conventional RS pipelines, whether collaborative filtering
\cite{koren2009matrix,rendle2010fm}, deep sequential models
\cite{kang2018sasrec,sun2019bert4rec}, or reinforcement-learning
optimizers \cite{mnih2015dqn,sutton2018rl}, operate
largely as \emph{single-agent systems}, optimizing one global objective
(e.g., click-through rate or watch time). This paradigm not only neglects competing goals, such as diversity, fairness, and
explainability \cite{zhang2020explain,burke2017fairrec},
but also hinders the system from adapting to the dynamic and complex nature of real-world environments, including heterogeneous content, evolving user intent, and complex feedback loops.
\cite{quadrana2018sequence,he2017ncf}.

Recent progress in multi-agent learning has introduced
decentralized and cooperative paradigms that decompose the
recommendation process into interacting roles.  
Each agent can specialize in tasks, such as perception, reasoning, or feedback
integration, jointly optimizing a shared objective through communication
and coordination \cite{wang2024generec,wang2025mavis}.  
These developments reveal that a multi-agent design can solve more
complex user problems, increasing recommendation quality and user
engagement \cite{boadana2025llmmusicagents}.

Concurrently, the emergence of foundation models (FMs) [large
language and multimodal models trained on vast corpora] has transformed
how recommender systems can represent, reason, and interact
\cite{vaswani2017attention,devlin2019bert,brown2020gpt3}.  
FMs enable zero-shot generalization \cite{he2023large,ranganathan2025zero}, natural-language interfaces, and
cross-modal reasoning over text, vision, and audio.  
When coupled with multi-agent coordination, they form the basis of
agentic recommender systems which autonomously plan, reflect, use tools and coordinate with other agents to achieve their goals.
\cite{he2020multimodule,wang2025mavis}.
 
Despite this rapid progress, the field lacks a unified taxonomy that bridges classical multi-agent reinforcement learning with these emerging foundation-model paradigms across diverse video ecosystems \cite{wu2023multimodalrec, zhang2021marlsurvey}. Prior surveys have focused either on Multi-Agent RL or on foundation models in traditional recommendation systems or collaboration in generic multi-agent systems, leaving a gap in understanding how these streams converge in modern recommender systems \cite{other2024survey}. Overall, this work aims to build that bridge for the domain of multi-agent video recommendation systems (MAVRS), outlining a
pathway toward self-improving, transparent, and trustworthy
next-generation video recommenders.\\

\noindent
\textbf{Why Video Recommenders}

\vspace{0.2cm}

\noindent
Although some of the
underlying principles presented in this paper can be generalized to other recommendation
domains,  the large-scale, high-impact nature of modern video recommenders makes them a perfect testbed for developing and validating LLM-powered multi-agent systems. While traditional architectures suffice for text or product IDs, video recommendation presents a unique 'modality gap' that necessitates agentic decomposition. Unlike text, which can be tokenized directly into an LLM's context window, video is high-dimensional, temporal, and multimodal. No single foundation model can currently ingest a user's entire long-term video watch history at the pixel level to perform reasoning. Multi-agent systems solve this by decoupling perception from reasoning: specialized 'Perception Agents' compress raw video into semantic summaries, while 'Reasoning Agents' utilize these lightweight textual representations to perform logic-heavy personalization. This modularity allows MAVRS to scale video understanding without hitting the context limits that plague single-model generative approaches.

\section{Background and Related Work}

Before the advent of multi-agent and LLM-driven frameworks, the field of recommender systems was dominated by two primary paradigms: collaborative filtering and content-based filtering. Collaborative filtering (CF) operates on the principle of homophily, identifying users with similar taste profiles to make recommendations based on what analogous users have enjoyed \cite{ricci2011recsys}.  Content-based (CB) methods, in contrast, focus on the intrinsic properties of items and recommend content with features similar to those a user has previously rated positively \cite{adomavicius2005,koren2009matrix}. While often effective, these classical approaches face challenges such as the "cold start" problem for new users or items, data sparsity in user-item interaction matrices, and a limited ability to capture the dynamic, multi-faceted nature of user intent \cite{burke2017fairrec}. These challenges paved the way for more complex, decentralized models, which form the basis of modern multi-agent systems \cite{quadrana2018sequence,he2017ncf,sun2019bert4rec,zhang2019deeplearning}. \\

\noindent
\textbf{Multi-Agent Recommender Systems}\\
\noindent
Early multi-agent recommender systems (MARS) emerged from distributed
AI research, where the goal was to decompose recommendation subtasks
among cooperative software entities \cite{wooldridge2009multiagent,selmi2014multi}.
Selmi \emph{et al.}~(2014)
identified four canonical roles: \emph{interface} agents that interact
with users, \emph{filtering} agents that match items to preferences,
\emph{learning} agents that update profiles, and \emph{mediator} agents
that resolve conflicts across heterogeneous sources.  Subsequent systems
incorporated negotiation, trust modeling, and content aggregation to
enhance autonomy and scalability \cite{burke2017fairrec}.  
Although these designs improved modularity, they relied heavily on
symbolic reasoning and rule-based communication, limiting adaptability
in large-scale, dynamic video environments.  
The success of deep reinforcement learning (DRL)---notably the
Deep Q-Network (DQN)~\cite{mnih2015dqn}---catalyzed a wave of research
towards optimizing multi-agent recommender systems using DRL
\cite{sutton2018rl,liebman2015djmc}.  
In MARL, multiple agents learn coordinated policies through shared or
partially shared rewards.  
Model-based methods such as MMRF optimize heterogeneous feedback signals
(e.g., watch-time, like-rate, dwell-time) using attention-based message
passing among agents for stable off-policy learning
\cite{wang2025mavis}. \\

\noindent
\textbf{Foundation-Model-Powered Recommendation}\\
\noindent
Foundation models (FMs),large language and multimodal
transformers, have redefined how recommender systems can represent and
reason about content \cite{vaswani2017attention,devlin2019bert,brown2020gpt3,radford2021learning,alayrac2022flamingo}.  
Large Language Models (LLMs) provide enhanced generalization abilities,
having trained on extensive datasets, allowing them to understand
complex patterns and handle new items or user trends effectively
\cite{chowdhery2023palm,touvron2023llama}.  
They offer improved explanation and reasoning capabilities by providing
more comprehensive and context-aware justifications
\cite{ouyang2022training,zhang2020explain}.  
Additionally, LLMs enhance personalization and interactivity through
their natural language processing features, enabling dynamic adaptation
to user feedback and preferences \cite{chen2024lifelong}.  
They can also allow users to have more fine-tuned control over the
system's understanding of user preferences and, subsequently, the
recommended content \cite{boadana2025llmmusicagents}.\\

LLMs have been integrated into RS through three main paradigms:
(i)~\emph{feature-based}, using FMs as embedding
extractors for user and item representations \cite{wang2024generec};
(ii)~\emph{generative}, treating recommendation as text or sequence
generation by prompting or fine-tuning \cite{brown2020gpt3,devlin2019bert};
and (iii)~\emph{agentic}, where the LLM serves as the core of autonomous
reasoning that plans, memorizes, and interacts through
natural language \cite{boadana2025llmmusicagents}. \\

\noindent
\textbf{Agentic Frameworks}\\
\noindent
Recent studies combine multi-agent coordination with LLM reasoning to
create conversational and collaborative recommenders.  
An LLM-based recommender agent is an autonomous entity designed to
perceive its environment, make decisions, and take actions within a
recommendation scenario \cite{chen2024lifelong}.  

MACRec and its extension MACRS organize LLM agents into hierarchical
roles---manager, analyst, searcher, reflector, and interpreter---to
perform sequential and dialogue-based tasks \cite{wang2024generec}.  
EmotionRec and MusicAgent further incorporate multimodal affect
detection, enabling personalized music and video recommendation grounded
in user emotion and context \cite{boadana2025llmmusicagents,yu2023musicagent}.  
These systems demonstrate that emotional awareness and cooperative
reasoning can significantly enhance engagement and trust
\cite{wang2025mavis}.

\begin{figure*}[hptb]
\centering
\includegraphics[width=1.5\columnwidth]{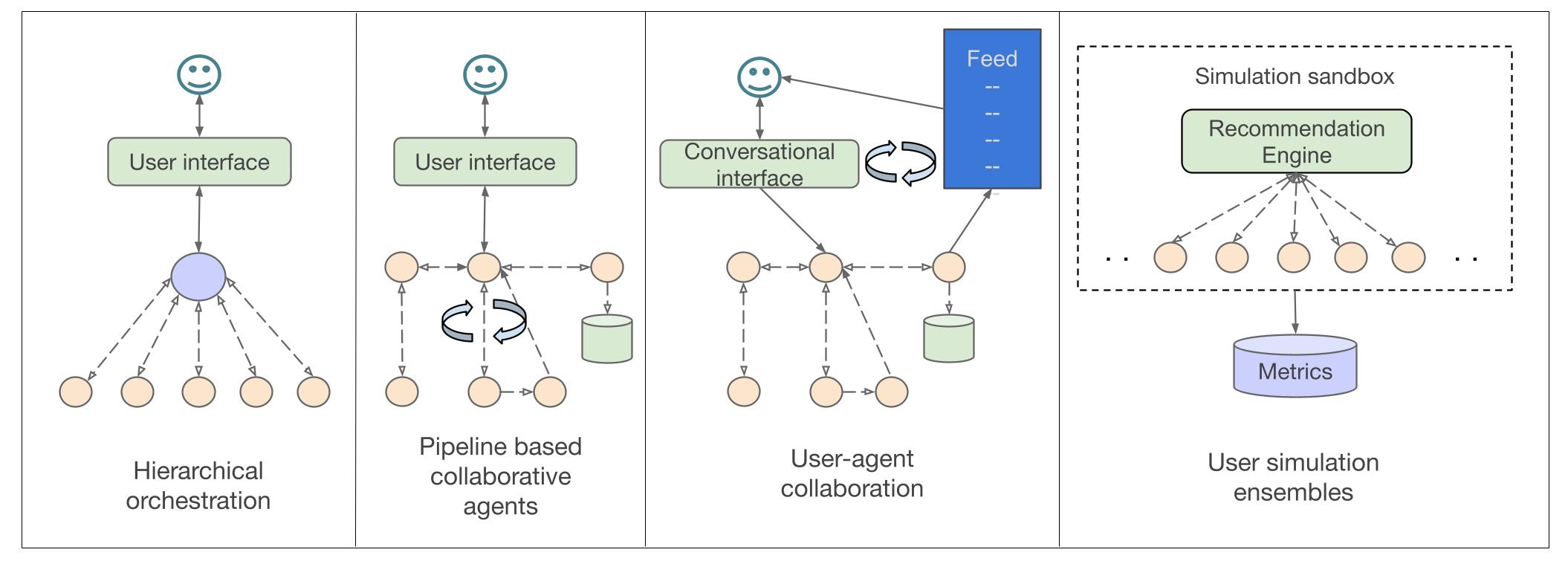} 
\caption{Illustration of Multi-agent Video Recommender patterns highlighting an example for each pattern in Section 3.}
\label{fig1}
\end{figure*}

\section{\textbf{Collaborative Multi-Agent Video Recommender Patterns}}

The collaborative interactions between LLM agents in video recommendation can be categorized into distinct architectural patterns. This taxonomy classifies systems according to the primary mechanism of agent interaction and the overarching goal of the collaboration, revealing how different structures are engineered to solve specific problems. The following sections detail prominent architectures, each illustrated with a key example from recent research \cite{wang2025mavis}.\\

\noindent
\textbf{3.1. Hierarchical Orchestration}\\

This architecture employs a central coordinating agent that directs the actions and integrates the outputs of specialized, subordinate agents to achieve a unified objective. The collaboration pattern is explicitly top-down, with the coordinating agent orchestrating the contributions of the agentic group. Subordinate agents may operate in two primary modes: (1) collaboratively, to jointly identify an optimal recommendation, or (2) \textit{competitively}, proposing distinct recommendations from which the coordinating agent selects based on user signals or other optimization criteria \cite{rahwan2019machine,wang2021dueling}.

A prominent example of this model is the \textbf{Model-based Multi-agent Ranking Framework (MMRF)}, \cite{zhou2024model} designed to maximize user WatchTime on a short-video platform. In MMRF, a main agent is dedicated to the primary objective (WatchTime) and is supported by auxiliary agents, each tasked with maximizing a secondary user interaction signal (e.g., Follow, Like, Comment). Coordination is achieved via an “Attentive Collaboration Mechanism,” which permits the main agent to dynamically weigh and integrate salient information from the auxiliary agents. This hierarchical structure allows the system to optimize for a primary metric while strategically leveraging correlated signals from secondary user preferences.

The \textbf{MMAgentRec} system \cite{xiao2025mmagentrec}, applied in the tourism domain, presents a conceptual variation. It prompts a single LLM to simulate multiple expert personas from diverse domains (e.g., natural sciences, social sciences, humanities), which then provide interdisciplinary advice on a user's request. This framework also incorporates a “reflection mechanism,” enabling the LLM to self-critique its outputs and refine its decision-making \cite{ouyang2022training}. This approach leverages the LLM's latent knowledge by structuring its reasoning process as an internal, collaborative dialogue among simulated experts \cite{boadana2025llmmusicagents}.

This architectural pattern can be generalized to multiple, distinct agents, each parameterized with specific prompts or inputs to optimize for different objectives. In a video RS context, this could be implemented as specialized agents recommending content from different domains (e.g., News, Education, Music) or optimizing for divergent engagement goals (e.g., long-term user value vs. short-term engagement) \cite{chen2023multiobjective,wang2025mavis}.\\

\noindent
\textbf{3.2. Pipeline-based Modular Collaboration}\\

In this architectural pattern, agents operate sequentially, forming a processing pipeline where each agent executes a distinct, specialized task. The output of one agent serves as the direct input for the next, establishing a modular workflow that decomposes a complex problem into manageable stages. This pattern is analogous to traditional, non-agentic industry systems where distinct engineering teams manage separate data processing pipelines (e.g., video processing and indexing, user history summarization, model training) that write intermediate outputs to offline databases \cite{other2024survey,adomavicius2005,he2017ncf,dasilva2023survey}.

The \textbf{VRAgent-R1} system demonstrates this approach, utilizing a two-stage pipeline to enhance video recommendation performance:

\begin{enumerate}
    \item \textbf{Item Perception (IP) Agent:} This initial agent processes raw, multimodal video content. It employs a “human-like progressive thinking” process to move beyond surface-level features, generating an enhanced semantic summary that captures latent, recommendation-relevant semantics \cite{radford2021learning,alayrac2022flamingo,li2023blip}.
    \item \textbf{User Simulation (US) Agent:} The semantic summary from the IP Agent enriches the base recommender model’s item representations. The US Agent leverages this enhanced understanding to simulate user decisions. This agent’s feedback is integrated into a reinforcement learning (RL) loop, with rewards for predicting the next video watched by the user and for providing Chain of Thought reasoning of whether a user would like a specific video. The resulting learned policy is better aligned with human preferences, and subsequently generates higher-quality recommendations \cite{mnih2015dqn,sutton2018rl}.
\end{enumerate}

In contrast to the two-stage VRAgent-R1, the authors of \textbf{MACRec} propose a conversational recommender system  with an alternative task decomposition \cite{wang2024macrec}:

\begin{itemize}
    \item \textbf{Manager:} Assigns sub-tasks to other agents, aggregates their responses, and reasons about the task status to generate a final response to the user or instantiate new sub-agents.
    \item \textbf{Reflector:} Evaluates the Manager’s proposed response and provides critical feedback for improvement. The Manager uses this feedback to decide whether to share the current recommendation with the user or iterate further \cite{ouyang2022training}.
    \item \textbf{User/Item Analyst:} Provides a nuanced analysis of both user preferences and item content. This role is analogous to the combined functions of the IP and US agents in VRAgent-R1.
    \item \textbf{Searcher:} Executes search queries and summarizes the results for the Manager. This two-stage process (search-then-summarize) optimizes token consumption for the Manager agent \cite{boadana2025llmmusicagents}.
    \item \textbf{Task Interpreter:} Interfaces with the user, converting natural language queries into structured task descriptions for the Manager. It also maintains the conversational state and history across multiple Manager calls \cite{fang2024macrs,huang2025fm4recsys}.
\end{itemize}

\noindent
\textbf{3.3. User-Agent Collaboration}\\

In this architecture, multiple agents collaborate internally to power a single, user-facing conversational interface (within a broader recommendation surface) where the primary objective is not to provide recommendations, but to empower the end-user with direct, intuitive control over their recommendation feed, thereby enhancing their “sense of agency” \cite{floridi2019ethicsai}.

\textbf{TKGPT} \cite{niu2025chat} is a system designed around this principle. It functions as an LLM-enhanced chatbot that allows users to modify their TikTok “For You” page through natural language. This is achieved through a partnership between two internal assistants. The \textbf{Recommender Assistant} interprets the user’s conversational requests to generate relevant keywords for video topics. The \textbf{Sorting Assistant} uses the LLM to assign weights to these keywords, which determine the \textit{proportion} of videos for each topic in the next batch of 32 videos. These videos are then shuffled and presented to the user.
This collaboration translates a user’s natural language intent into concrete algorithmic adjustments via a proportional allocation and batch-based update mechanism, creating a direct and transparent control interface \cite{huang2025fm4recsys,fang2024macrs}.\\

\noindent
\textbf{3.4. User Simulation Agent Ensembles}\\

This architecture uses agents not as the core recommender, but as a simulated population of users. The goal is to generate high-fidelity synthetic interaction data, which can be used to evaluate system performance offline, train other models, or study complex user behavior phenomena without the cost and risk of live A/B testing \cite{wang2025mavis,rahwan2019machine}.

\textbf{Agent4Rec} \cite{zhang2024generative} is the primary example of this pattern, creating a simulator with thousands of LLM-empowered generative agents \cite{wang2025mavis}. Each agent is initialized from real-world datasets with a detailed profile, including unique tastes and social traits like activity (interaction frequency) and conformity (alignment with popular sentiment). The central goal is to achieve “agent alignment” by ensuring simulated behaviors are faithful to those of real humans, allowing the ensemble to replicate effects like the “filter bubble” \cite{zhang2024generativeagents}. The \textbf{US Agent} from VRAgent-R1 also serves as a simulation agent. These two systems exemplify different philosophies for achieving alignment: Agent4Rec relies on rich, static profiling initialized from real data, whereas VRAgent-R1’s US Agent uses a dynamic, in-loop training method—Reinforcement Learning with Group Relative Policy Optimization (GRPO)—to continuously align its behavior with real user decisions \cite{chen2025vragentr1}.

This simulation pattern can be used to create a sandbox for testing multi-agent systems' insights on social norms and governance. For example, Agent4Rec's modeling of user ensembles allows researchers to prototype various agent incentive formulations and observe emergent behaviors (like filter bubbles) without real-world risk.

\begin{table*}[t]
\centering
\caption{Evaluation of collaborative multi-agent video recommender architectures. Metrics emphasize coordination, user alignment, and computational feasibility.}
\label{tab:evaluation}
\scriptsize
\setlength{\tabcolsep}{6pt}
\renewcommand{\arraystretch}{1.0}
\begin{tabularx}{\textwidth}{L{0.16\textwidth} L{0.24\textwidth} L{0.23\textwidth} L{0.27\textwidth}}
\toprule
\textbf{Pattern} & \textbf{Primary Evaluation Focus} & \textbf{Representative Metrics} & \textbf{Critical Failure Points \& Risks} \\
\midrule
\textbf{Hierarchical Orchestration}     (e.g., MMRF, MMAgentRec) & \textbf{Orchestration Effectiveness:} How well does the central agent integrate diverse sub-goals to optimize the primary system objective? &
Main objective metric (e.g., WatchTime), contribution weights (from attentive mechanism), system-wide latency. &
\textbf{Coordinator Bottleneck:} The central agent becomes a single point of failure. \textbf{Conflicting Goals:} Auxiliary agents may work at cross-purposes, harming the main objective.\\
\midrule
\textbf{Pipeline-based Modular} (e.g., VRAgent-R1, MACRec) &
\textbf{End-to-End Task Quality:} How well does the final output perform after passing through all sequential stages? &
Quality of intermediate outputs, error propagation rate, end-to-end latency. &
\textbf{Compounding Errors and Brittleness:} An error in an early agent (e.g., IP Agent) can degrade the entire chain.\\
\midrule
\textbf{User-Agent Collaboration} (e.g., TKGPT) &
\textbf{User-Perceived Agency:} Does the user feel in control and satisfied with the system’s response to their natural language commands? &
User satisfaction (SUS scores), task success rate (from user studies), latency from command to feed update. &
\textbf{Misinterpretation:} The system may misunderstand the user’s (often ambiguous) intent and make drastic, undesirable changes to recommendations.\\
\midrule
\textbf{User Simulation Ensemble} (e.g., Agent4Rec) &
\textbf{Behavioral Fidelity:} How accurately does the simulated agent population replicate the statistical properties of real human users? &
KL divergence (or similar) between simulated and real interaction distributions; replication of known macro-effects (e.g., filter bubbles). &
\textbf{Lack of Generalization:} Agents overfit to initialization data and fail to model novel behaviors.

\textbf{Prohibitive Cost:} High computational overhead for running thousands of LLM agents.\\
\bottomrule
\end{tabularx}
\end{table*}

\section{Agent-centric Evaluation}

Evaluating multi-agent recommender systems (MARS) differs fundamentally from classical single-model recommenders because multiple agents interact, negotiate, and learn concurrently \cite{dafoe2021cooperative,zhang2024generativeagents}. Standard metrics such as Precision@K and NDCG remain necessary to measure the quality of the recommendations \cite{adomavicius2005,he2017ncf} but are insufficient to capture coordination, reasoning quality, and emergent behaviors of the agentic framework itself \cite{huang2025fm4recsys,zhang2024trust}. A comprehensive evaluation must therefore be multi-dimensional, assessing not only the final output but also the internal processes of the agents \cite{other2024survey,zhang2024generativeagents}. We propose five key dimensions for a holistic, agent-centric evaluation.\\

\noindent
\textbf{4.1. Task-Specific Quality}\\
\noindent
This dimension evaluates the performance of an individual agent on its specialized sub-task, separate from the final recommendation \cite{zhang2024generativeagents,ouyang2022training}.

\begin{itemize}
    \item \textbf{For Perception Agents} (e.g., the IP Agent in VRAgent-R1): Evaluation can involve comparing the agent-generated representation/summary for a sample of videos against human-generated summaries or ground-truth labels using metrics like ROUGE, BERTScore, or emotion-based recognition signals \cite{radford2021learning,alayrac2022flamingo,li2023blip,chaugule2016facialreview}.
    \item \textbf{For Reasoning Agents} (e.g., the "reflection mechanism" in MMAgentRec): Evaluation is often qualitative, assessing the logical coherence, factuality, and self-correction capability of the agent's internal monologue or "scratchpad" \cite{ouyang2022training}.
    \item \textbf{For Specialized Recommenders} (e.g., the auxiliary agents in MMRF): These can be evaluated on their own proxy metrics (e.g., can the 'Like' agent predict 'Likes' with high precision?).
\end{itemize}

\noindent
\textbf{4.2. Coordination \& Collaboration Efficiency}\\
\noindent
This dimension assesses the \textit{interaction} between agents, focusing on the overhead and effectiveness of their collaboration.

\begin{itemize}
    \item \textbf{Communication Overhead:} This is a critical metric for LLM-based systems, measured in the number of tokens, messages, or API calls exchanged between agents to reach a decision. The "Searcher" agent in MACRec is an example of a design that explicitly optimizes this \cite{huang2025fm4recsys,zhang2024generativeagents}.
    \item \textbf{Latency:} The end-to-end time from user request to final recommendation. This is vital for real-time video feeds and includes the cumulative processing and communication time of all agents in the chain \cite{dafoe2021cooperative,huang2025llmars}.
    \item \textbf{Contribution Alignment:} In hierarchical systems like MMRF, this measures whether the auxiliary agents' contributions (e.g., 'Follow' signal) are weighted appropriately and genuinely improve the main agent's primary objective ('WatchTime').
\end{itemize}

\noindent
\textbf{4.3. System-Level \& Emergent Properties}\\
This dimension evaluates the macro-behavior of the entire system, particularly its stability and adaptability \cite{zhang2024generativeagents,fang2024macrs}.

\begin{itemize}
    \item \textbf{Robustness \& Fault Tolerance:} This tests how the system handles the failure of a single agent. Does a pipeline-based system collapse (a "brittle" failure), or can a hierarchical system's coordinator route around the failed agent \cite{fang2024macrs,dafoe2021cooperative}?
    \item \textbf{Adaptability:} This measures how quickly the agent ensemble can adapt to new items, new user interests, or a shift in the data distribution. This is a key goal for systems using RL (like VRAgent-R1) and "lifelong personalization" \cite{mnih2015dqn,sutton2018rl,chen2024lifelong}.
    \item \textbf{Emergent Behavior Accuracy:} For user simulation ensembles like Agent4Rec, this is the primary evaluation. It involves measuring the statistical divergence (e.g., KL divergence) between the simulated interaction data and real user data \cite{zhang2024generativeagents,zhang2024trust}.
\end{itemize}

\noindent
\textbf{4.4. Human-Alignment \& User-Centric Metrics}\\
\noindent
This dimension moves beyond offline metrics to measure the system's impact on the end-user experience, which is often the primary goal \cite{zhang2024trust,zhang2024generativeagents,dafoe2021cooperative,chaugule2016facialreview}.

\begin{itemize}
    \item \textbf{Controllability \& Agency:} For systems like TKGPT, the core metric is the user's "sense of agency." This is measured via user studies, assessing whether users feel their natural language commands are correctly interpreted and lead to a satisfying change in their feed \cite{zhang2024trust,zhang2024generativeagents}.
    \item \textbf{Explainability:} A MARS architecture should naturally provide better explainability \cite{zhang2020explain,zhang2019deeplearning}. Evaluation can involve user studies where participants rate the quality of explanations generated by the system (e.g., "The 'Education' agent suggested this video, and the 'Sorting' agent prioritized it because you asked for 'deep dives'") \cite{zhang2024trust,dafoe2021cooperative}.
    \item \textbf{Trustworthiness:} This is a longitudinal user-study metric measuring whether users trust the system's recommendations and explanations over time \cite{zhang2024generativeagents,floridi2019ethicsai}.
    \item \textbf{Fairness:} The quality of reasoning agents and user simulation agents strongly affects bias in the recommendations for specific slices or users or content \cite{burke2017fairrec,mehrabi2021fairness}. Standard fairness metrics that measure equal exposure for items, such as Jain's Index or Gini Index, and metrics based on user group disparity (like Equalized Odds or Demographic Parity) can be used to measure end-to-end fairness \cite{wang2023survey,zhang2024trust}.\\
\end{itemize}

\noindent
\textbf{4.5. Scalability \& Economic Viability}\\
\noindent
This practical dimension assesses the cost of deploying and maintaining the MARS \cite{shleifer2023costs,chen2024efficient,zhang2024generativeagents}. For LLM-driven agents, the total token cost per user request or per recommendation batch and the end-to-end latency for the coordinating agents to generate a recommendation \cite{shleifer2023costs,zhang2024generativeagents} are important to measure. For systems using RL (VRAgent-R1) or large-scale simulation (Agent4Rec), the computational resources (GPU hours, real-user data) required to train or align the agents before they produce high-fidelity results \cite{wu2024economicsagents,zhang2024generativeagents} can be measured.

\section{Challenges and Open Problems}

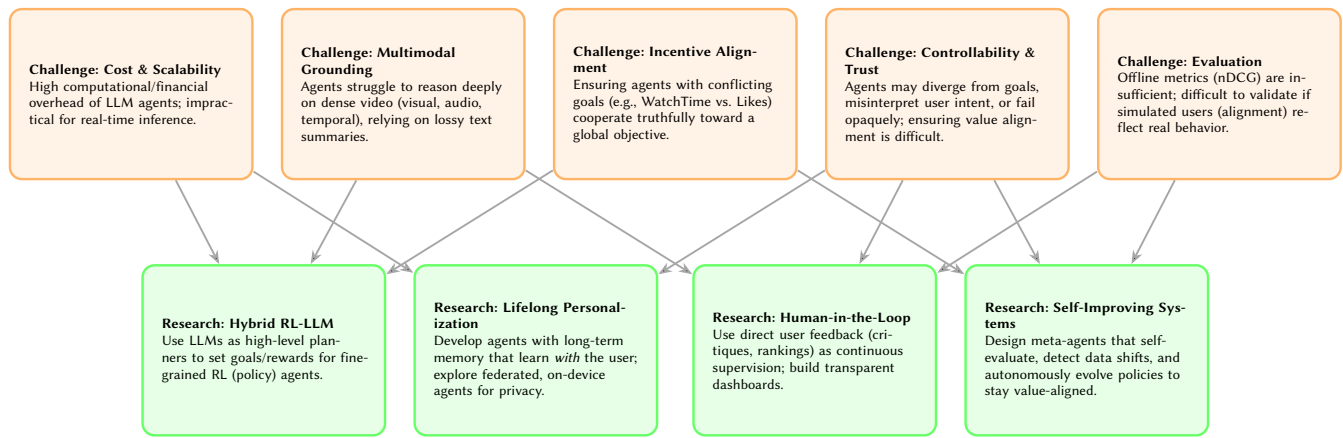
\begin{figure*}[ht]
    \centering
    \resizebox{1.0\textwidth}{!}{%
    \begin{tikzpicture}[
        node distance=0.4cm, 
        every node/.style={font=\sffamily\scriptsize} 
    ]
    
    \tikzset{
      base/.style = {
        rectangle, 
        rounded corners=5pt, 
        draw, 
        line width=1pt, 
        text width=3 cm, 
        align=left, 
        inner sep=8pt,
        minimum height=2.5cm
      },
      start/.style = {
        base, 
        fill=blue!10, 
        draw=blue!60, 
        text width=7cm, 
        align=center,
        minimum height=0pt
      },
      challenge/.style = {
        base, 
        fill=orange!10, 
        draw=orange!60
      },
      research/.style = {
        base, 
        fill=green!10, 
        draw=green!60
      },
      arrow/.style = {
        -{Stealth[length=2mm, width=1.5mm]}, 
        line width=0.8pt, 
        draw=gray!70
      }
    }

    

     \node (c5) [challenge] {
        {\textbf{Challenge: Incentive Alignment}}\\
        Ensuring agents with conflicting goals (e.g., WatchTime vs. Likes)
        cooperate truthfully toward a global objective.
    };
    
    \node (c2) [challenge, left=of c5] {
        {\textbf{Challenge: Multimodal Grounding}}\\
        Agents struggle to reason deeply on dense video
        (visual, audio, temporal), relying on lossy text summaries.
    };
    \node (c1) [challenge, left=of c2] {
        {\textbf{Challenge: Cost \& Scalability}}\\
        High computational/financial overhead of LLM agents;
        impractical for real-time inference.
    };
    \node (c4) [challenge, right=of c5] {
        {\textbf{Challenge: Controllability \& Trust}}\\
        Agents may diverge from goals, misinterpret user intent,
        or fail opaquely; ensuring value alignment is difficult.
    };
    \node (c3) [challenge, right=of c4] {
        {\textbf{Challenge: Evaluation}}\\
        Offline metrics (nDCG) are insufficient; difficult to validate
        if simulated users (alignment) reflect real behavior.
    };
    \node (r_anchor) [coordinate, below=2.5cm of c5] {};
    
    \node (r2) [research, left=0.25cm of r_anchor] {
        {\textbf{Research: Lifelong Personalization}}\\
        Develop agents with long-term memory that learn \emph{with} the user;
        explore federated, on-device agents for privacy.
    };
    \node (r3) [research, right=0.25cm of r_anchor] {
        {\textbf{Research: Human-in-the-Loop}}\\
        Use direct user feedback (critiques, rankings) as
        continuous supervision; build transparent dashboards.
    };
    \node (r1) [research, left=of r2] {
        {\textbf{Research: Hybrid RL-LLM}}\\
        Use LLMs as high-level planners to set goals/rewards
        for fine-grained RL (policy) agents.
    };
    \node (r4) [research, right=of r3] {
        {\textbf{Research: Self-Improving Systems}}\\
        Design meta-agents that self-evaluate, detect data shifts,
        and autonomously evolve policies to stay value-aligned.
    };

    

    \path [arrow] (c1) edge (r1)
                       edge (r2);
    
    \path [arrow] (c2) edge (r1)
                       edge (r3);
                       
    \path [arrow] (c3) edge (r3)
                       edge (r4);

    \path [arrow] (c4) edge (r2)
                       edge (r3)
                       edge (r4);
                       
    \path [arrow] (c5) edge (r1)
                       edge (r4);

    \end{tikzpicture}
    }
    \caption{ Challenges and Future Research Directions for Multi-Agent Video Recommendation Systems (MAVRS).}
    \label{fig:mavrs_challenges}

\end{figure*}

Despite the rapid progress in LLM-powered multi-agent recommenders, deploying MAVRS at industry scale presents significant challenges, limiting their current utility and trustworthiness \cite{other2024survey,chen2025vragentr1}.\\

\noindent
\textbf{5.1 Computational Cost and Scalability} \\
\noindent
The reliance on large language models (LLMs) as the cognitive core for agents introduces significant computational and financial overhead. Architectures like Agent4Rec, which simulate \textit{thousands} of agents, are prohibitively expensive for most research labs and impractical for real-time training or inference in production RS \cite{shleifer2023costs}. Lightweight, "distilled" agent models or more efficient token-sharing mechanisms might offer a path forward to widespread adoption \cite{chen2024efficient,other2024survey}.\\

\noindent
\textbf{5.2 Multimodal Grounding and Reasoning} \\
\noindent
Video is an inherently dense medium packed with informationa cross modalities: visual, audio, textual and temporal. Current agents, especially those built on text-centric LLMs, struggle to "ground" their reasoning in this rich data. While systems like VRAgent-R1 employ an \textit{Item Perception (IP) Agent} to generate semantic summaries, this is often a lossy compression \cite{li2023blip}. The challenge lies in enabling agents to perform deep, cross-modal reasoning directly on video streams, moving beyond metadata and text summaries to cohesively understanding the content of the video \cite{alayrac2022flamingo,radford2021learning,he2023large,huang2025fm4recsys}.\\

\noindent
\textbf{5.3 Evaluation}\\
\noindent
As discussed in the previous section, evaluating the performance of complex, collaborative agent systems is an open problem. Offline metrics (e.g., nDCG, MRR) may not capture the subjective benefits of context-aware, conversational recommendation \cite{adomavicius2005,he2017ncf}. Furthermore, user simulation ensembles (Agent4Rec, VRAgent-R1) face an alignment problem: ensuring that synthetic agent behavior is a high-fidelity proxy for real human behavior, including irrationality, conformity, and drift \cite{chen2025vragentr1,rahwan2019machine}. Without robust validation, it is difficult to trust simulation-based findings or offline training \cite{zhang2024trust}.\\

\noindent
\textbf{5.4 Controllability and Trustworthiness} \\
\noindent
As agents become more autonomous, ensuring they are controllable, robust, and aligned with human values becomes essential \cite{floridi2019ethicsai,zhang2024trust,huang2025fm4recsys}. In hierarchical systems (MMRF), a subordinate agent could diverge and optimize its secondary metric at the expense of the primary goal \cite{chen2025vragentr1}. In conversational systems (TKGPT), the translation of user intent into algorithmic action must be transparent and faithful \cite{li2023blip,zhang2020explain}. Agents could also fail in a silent, opaque manner, causing errors to propagate through other downstream agents \cite{rahwan2019machine,ouyang2022training}.\\

\noindent
\textbf{5.5 Incentive Alignment}\\
\noindent
In multi-agent systems, agents must be \textit{incentivized} to collaborate effectively \cite{rahwan2019machine}. In current recommenders, this is implicit (e.g., optimizing a shared goal). However, as systems grow in complexity, agents with different objectives (e.g., user WatchTime vs. user Likes in MMRF) may enter into conflict. A key challenge is to design explicit coordination mechanisms, potentially borrowing from computational economics (e.g., auctions, contract theory) \cite{zhang2024generativeagents,ostrom1990governing}. These mechanisms can help the high-level agent ensure subordinate agents cooperate truthfully and robustly toward the global system objective, even under uncertainty or conflicting signals \cite{huang2025fm4recsys,zhang2024trust}. However, unlike computational economics, incentives in LLM-based agents are configured via natural language, which allows room for the underlying LLM to interpret the prompt in ways that differ from what the developer intended. \cite{yang2020learning}\\

\vspace{-0.75cm}
\section{Future Directions}

Addressing the challenges above requires unifying algorithmic efficiency, realistic evaluation, and human alignment \cite{fang2024macrs,zhang2024trust}.  
Future research should treat multi-agent recommendation as a socio-technical system integrating cognition, collaboration, and ethics \cite{floridi2019ethicsai,rahwan2019machine}. These challenges also highlight specific directions for future research, focusing on the development of more intelligent, adaptive, and human-centric systems.\\

\noindent
\textbf{6.1. Hybrid RL-LLM Architectures} \\
\noindent
A promising frontier is the deeper integration of Reinforcement Learning (RL) and LLMs.  
LLMs excel at high-level reasoning, planning, and understanding user intent (as seen in TKGPT or the \textit{Manager} MACRec), while RL excels at fine-grained policy optimization in dynamic environments (as seen in VRAgent-R1).  
Future systems may use an LLM as a "planner" to set high-level goals or generate reward-shaping functions for a subordinate RL agent, creating a hybrid system that is both context-aware and adaptive to user feedback \cite{sutton2018rl,mnih2015dqn,zhang2024generativeagents}.  
These emerging “planner–executor” hybrid systems show promise for scaling such coordination while maintaining explainablity \cite{garnelo2019neuralsymbolic,li2023blip}.\\

\noindent
\textbf{6.2. Lifelong Personalization and Agent Memory} \\
\noindent
Current models largely operate on a session- or user-profile-level memory.  
The next step is lifelong personalization, where agents build and maintain a dynamic, long-term memory of user preferences and evolving interests.  
This involves moving beyond static profiles (Agent4Rec) to models where agents can reason over their interaction history, self-correct past assumptions, and proactively adapt to a user's long-term personal journey, effectively \textit{learning with} the user.  
This requires new designs for maintaining a summarized version of long-term user preference history \cite{chen2024lifelong,p4lm2024,rec-r12024}.  

A promising research area here is Federated Collaboration, which applies federated learning principles to the multi-agent paradigm.  
A local "User Profile Agent," co-located with the user (such as on the device), could perform deep, lifelong personalization using raw interaction data that never leaves the device.  
The local agent can interact with online RS agents while optimizing for privacy and user well-being \cite{shleifer2023costs,huang2025fm4recsys}.\\

\noindent
\textbf{6.3. Human-in-the-Loop Validation} \\
\noindent
Long-term trust depends on user participation \cite{burke2017fairrec,zhang2020explain}.  
Crowdsourced or platform-integrated feedback, where users critique and rank recommendations, can serve as continuous supervision \cite{huang2025fm4recsys,fang2024macrs}.  
Interactive dashboards visualizing reasoning and fairness trade-offs will enhance transparency and literacy among users and regulators \cite{zhang2024trust,floridi2019ethicsai}.  
In the long term, we can derive these signals directly using optimized multimodal affect detection (e.g., facial expression or tone analysis) to enhance personalization \cite{chaugule2016facialreview}.\\

\noindent
\textbf{6.4. Toward Self-Improving Recommenders}\\
\noindent
The next frontier is self-governing ecosystems where agents perceive, reason, and evolve collaboratively \cite{fang2024macrs,wang2025mavis}.  
Such multi-agent architectures should enable a meta-agent to evaluate reasoning quality, detect distributional shifts, and autonomously propose schema or policy updates \cite{chen2024lifelong,huang2025fm4recsys}.  
The system should understand cause and effect and evolve its strategies to achieve better outcomes than optimizing for short-term objectives like watch time \cite{peters2017elements,scholkopf2021toward}.  
By self-reflecting to continuously optimizing the behavior and incentives of the modular internal agents, these multi-agent systems can evolve from content delivery tools into recommenders that are closely aligned with human values \cite{floridi2019ethicsai,zhang2024trust}.\\

\bibliographystyle{ACM-Reference-Format}
\bibliography{main}


\end{document}